\documentclass[10pt,conference]{IEEEtran}
\IEEEoverridecommandlockouts
\usepackage{bbm}
\usepackage{amsmath}
\usepackage{amsfonts}
\usepackage[dvips]{graphicx}
\usepackage{times}
\usepackage{cite}
\usepackage{array}
\usepackage{mathrsfs}
\usepackage{multirow}
\usepackage{amssymb}

\usepackage{stfloats}
\usepackage{subfigure}
\usepackage{footnote}
\usepackage{booktabs}
\usepackage{array}
\usepackage{algorithm}
\usepackage{subeqnarray}
\usepackage{cases}
\usepackage{threeparttable}
\usepackage{xcolor}
\usepackage{hyperref}
\usepackage{epstopdf}
\usepackage{algpseudocode}
\usepackage{bm}
\usepackage{multirow}
\usepackage{adjustbox}
\usepackage{stfloats}

\def\BibTeX{{\rm B\kern-.05em{\sc i\kern-.025em b}\kern-.08em
		T\kern-.1667em\lower.7ex\hbox{E}\kern-.125emX}}

\begin{document}
	
\title{Deep Learning for Hierarchical Beam Alignment in mmWave Communication Systems}
\author{Junyi Yang, Weifeng Zhu, and Meixia Tao\\
Department of Electronic Engineering, Shanghai Jiao Tong University, Shanghai, China \\
Emails: \{yangjunyi, wf.zhu, mxtao\}@sjtu.edu.cn


}

\maketitle	
\begin{abstract}
Fast and precise beam alignment is crucial to support high-quality data transmission in millimeter wave (mmWave) communication systems. In this work, we propose a novel deep learning based hierarchical beam alignment method that learns two tiers of probing codebooks (PCs) and uses their measurements to predict the optimal beam in a coarse-to-fine searching manner. Specifically, the proposed method first performs coarse channel measurement using the tier-1 PC, then selects a tier-2 PC for fine channel measurement, and finally predicts the optimal beam based on both coarse and fine measurements. The proposed deep neural network (DNN) architecture is trained in two steps. First, the tier-1 PC and the tier-2 PC selector are trained jointly.  After that, all the tier-2 PCs together with the optimal beam predictors are trained jointly. The learned hierarchical PCs can capture the features of propagation environment. Numerical results based on realistic ray-tracing datasets demonstrate that the proposed method is superior to the state-of-art beam alignment methods in both alignment accuracy and sweeping overhead.

\end{abstract}

\section{Introduction}

Millimeter wave (mmWave) communication plays a key role in 5G/6G technologies for its wide spectrum resource located between 28GHz and 300 GHz \cite{background1_hierarchical1}.
Compared with the conventional  sub-6GHz counterpart, communication at the mmWave band suffers from harsher propagation conditions. Directional beamforming using large-scale antenna arrays at the base station (BS) and the user equipment (UE) is often employed to compensate the severe path loss\cite{background2}.
In practice, to ensure high-quality data transmission in mmWave communication, beam alignment is needed to find the best directional beamformer in a pre-defined codebook.
Since the codebook usually consists of a large number of narrow beams and the mmWave communication is very sensitive to the dynamic environment, it is crucial to design high-accuracy and low-overhead beam alignment strategies.



A conventional approach for beam alignment is to perform an exhaustive search of the candidate beams in the codebook.
However, the exhaustive search method usually suffers from high time cost on the brute-force beam sweeping and therefore results in undesirable latency and signaling overhead, which limits its practical usage \cite{exhaustive1}\cite{exhaustive2}.
Hierarchical beam search is an alternative approach based on beam sweeping, which can reduce the signaling overhead by utilizing a multiple-tier codebook. Under the hierarchical beam search framework, the BS usually sweeps several wider beams and then gradually focuses on a thinner search space to seek the optimal beam. Note that a well-designed codebook can effectively improve the performance of the optimal beam search  \cite{hierarchical2}\cite{hierarchical3}.
However, the performance of the hierarchical search is sensitive to the wide beams with imperfect patterns and noise, which can easily result in error accumulation in the searching process.

Recently, deep learning (DL) is regarded as a promising technology to improve both the accuracy and the signaling overhead of beam alignment.
In the works \cite{ML1,ML2,ML3_rosslyn}, several DL methods are proposed to find the optimal BS and the optimal beam to serve the UE based on its location information. However, acquiring the location information of UEs needs additional sensors and additional feedback.
On the other hand, deep reinforcement learning (DRL) methods are also proposed to gradually find the optimal beam by performing repeated interactions between the BS and the UE \cite{reinforcement1,reinforcement2,reinforcement3}. The frequent interactions may cause larger latency and more control signal.
To capture the characteristics of the propagation environment and the channel information effectively, the work \cite{Jeffrey} proposes to learn the probing codebook (PC) along with the beam predictor. Note that the learned PC can only provide limited performance when the codebook size is small.



In this paper, by leveraging DL, we propose a deep neural network (DNN)-based mmWave hierarchical beam alignment method.
Compared with the method in \cite{Jeffrey}, the BS adopts a 2-tier PC to probe more comprehensive channel information with the fixed sweeping overhead.
Our contributions can be summarized as follows:
\begin{enumerate}
\item We propose a DL-based hierarchical search method for beam alignment, where the BS employs two tiers of learnable PCs to predict the optimal beam in a coarse-to-fine searching manner. Similar to \cite{Jeffrey}, we use complex neural network (NN) layers to model PCs. But the application of the hierarchical structure enriches the PC space to achieve a significant performance improvement. In addition, we introduce a selector and multiple beam predictors to assist with the beam search.
\item An effective training strategy is proposed for our DNN. Unlike \cite{Jeffrey} which performs a single-step training directly, we perform a two-step training for the DNN to avoid overfitting. In the first step, we propose a clustering-based method to generate labels, then train the tier-1 PC and the selector jointly for coarse channel measurement. After the training in the first step is finished, we train the tier-2 PCs and beam predictors jointly for fine channel measurement.
\item We simulate the proposed method in ray-tracing datasets of outdoor and indoor environments which are generated by Wireless InSite \cite{software}.
The numerical results show that in both environments our method can consistently achieve a 5\%$\sim$14\% higher beam alignment accuracy comparing to the method proposed in \cite{Jeffrey} with the same sweeping overhead.
In addition, in outdoor environment at 28GHz, the proposed method can achieve a similar performance but with only 60.5\% sweeping overhead compared with the conventional 2-tier hierarchical beam search. Results also show that in indoor environment at 60GHz, the proposed method is significantly superior to the conventional 2-tier hierarchical beam search, and can even outperform the exhaustive search.

\end{enumerate}


\section{System Model}

We consider a downlink multiple-input single-output (MISO) system consisting of one BS and one UE, where the BS is equipped with $N_t$ antennas and the UE has only one antenna. In addition, the scenario is dynamic and the UE is assumed to be mobile.
Here, we concentrate on the beam alignment at the BS side. The beam alignment at the UE side can be similarly performed if there are multiple antennas at the UE.
For simplicity, we assume that the uniform linear array (ULA) is equipped at the BS. Note that our proposed method can also be extended to the other array geometries.
We only consider the radio frequency (RF) domain analog beamforming in the BS for the purpose of beam alignment. Each antenna element is connected to a dedicated analog phase shifter thus the analog beamformer is given as
\begin{equation}\label{equ3}
\mathbf{v} = \frac{1}{\sqrt{N_t}}\left[e^{\mathrm{j}\phi_{1}},e^{\mathrm{j}\phi_{2}} ,\dots,e^{\mathrm{j}\phi_{N_t}}\right]^T,
\end{equation}
where $\phi_i$ represents the phase of the $i$th element.
Here, we adopt the common discrete fourier transform (DFT) codebook $\mathbf{V} = [\mathbf{v}_1, \mathbf{v}_2, \dots,\mathbf{v}_{N_{\mathbf{V}}}] \in \mathbb{C}^{N_t \times N_{\mathbf{V}}}$ for downlink transmission. In the DFT codebook, each beam steers to a discrete direction and the beam codeword $\mathbf{v}_{i}$ is given by

\begin{equation}
\begin{aligned}
    \mathbf{v}_{i} = \frac{1}{\sqrt{N_{t}}} \left[ 1,e^{\omega_i},\dots,e^{\mathrm{j}(N_t-1)\omega_i} \right]^T,
\end{aligned}
\end{equation}
where $\omega_i =\frac{2\pi d}{\lambda}\frac{(2(i-1)-N_{\mathbf{V}})}{N_{\mathbf{V}}}$, $\lambda$ represents the carrier wavelength and $d$ is the antenna spacing. The codebook size $N_{\mathbf{V}}$ is usually assumed to be large enough to cover the whole space. When a symbol $s \in \mathbb{C}$ with unit power constraint $\mathbb{E}(|s|^2) = 1$ is transmitted using beam $\mathbf{v}_{i}$, the received signal $y$ at the UE can be expressed as
\begin{equation}\label{equ4}
y = \sqrt{\rho}\mathbf{h}^H\mathbf{v}_{i}s + n,
\end{equation}
where $\mathbf{h}\in\mathbb{C}^{N_t\times1}$ is the narrowband MISO mmWave channel between the BS and the UE, $\rho$ is the transmit power and $n$ is the complex Gaussian noise with zero mean and variance $\sigma^2_{n}$. The received signal-to-noise ratio (SNR) at the UE with channel $\mathbf{h}$ using beam $\mathbf{v}_{i}$ is
$\text{SNR} = \frac{\rho|\mathbf{h}^H \mathbf{v}_{i}|^2}{\sigma^2_{n}}$.

For the given DFT codebook $\mathbf{V}$, our target is to find the optimal beam codeword that realizes the maximal SNR for transmission:
\begin{align}\label{equ6}
i^*_{\mathbf{V}} =\mathop{\arg\max}\limits_{i \in \{ 1,\dots,N_\mathbf{V} \} } \left(\frac{\rho|\mathbf{h}^H\mathbf{v}_i|^2}{\sigma^2_{n}}\right) = \mathop{\arg\max}\limits_{i \in \{ 1,\dots,N_\mathbf{V} \} }(|\mathbf{h}^H\mathbf{v}_i|^2).
\end{align}


\section{The Proposed Method}

The proposed beam alignment method falls into the beam sweeping framework, where the BS sweeps a PC and then finds the optimal beam in the DFT codebook $\mathbf{V}$ with the probed channel information. To obtain a comprehensive measurement of the channel, the PC has a hierarchical structure, where the first tier contains one coarse-search codebook and the second tier contains $G$ fine-search codebooks. Let $\mathbf{W}^{\rm{c}} = [\mathbf{w}^{\rm{c}}_{1}, \dots, \mathbf{w}^{\rm{c}}_{N_1}] \in \mathbb{C}^{N_t \times N_1}$ denote the coarse-search codebook with $\mathbf{w}^{\rm{c}}_{i}$ being the $i$th beam in the codebook and $N_1$ representing the codebook size. Likewise, let $\{\mathbf{W}^{\rm{f}}_1,\dots,\mathbf{W}^{\rm{f}}_G\}$ denote the set of fine-search codebooks with each $\mathbf{W}^{\rm{f}}_k = [\mathbf{w}^{\rm{f}}_{k,1},\dots,\mathbf{w}^{\rm{f}}_{k,N_2}] \in \mathbb{C}^{N_t \times N_2}$, where $k \in \{1,2,\dots,G\}$ is the index of the fine-search codebook. Here, we assume the $G$ fine-search codebooks have the same size $N_2$. The size of the PC is usually much smaller than the DFT codebook size $N_{\mathbf{V}}$. With the two-tier PC, the proposed method performs beam alignment in two steps.


In the first step, the BS sweeps the coarse-search codebook $\mathbf{W}^{\rm{c}}$. The UE measures the received power, then reports the measurement to the BS, which is given as
\begin{equation}\label{equ:m1}
    \mathbf{z}^{\rm{c}} = \left[|y^{\rm{c}}_{1}|^2, \dots, |y^{\rm{c}}_{N_1}|^2\right]^T,
\end{equation}
where $y^{\rm{c}}_{i} = \sqrt{\rho}\mathbf{h}^H\mathbf{w}^{\rm{c}}_{i} s + n^{\rm{c}}$ is the received signal of the beam $\mathbf{w}^{\rm{c}}_{i}$ in the UE. Based on the reported $\mathbf{z^{\rm{c}}}$, the BS utilizes the selector $f(\cdot)$ to select one of the $G$ fine-search codebooks, denoted as $\mathbf{W}^{\rm{f}}_{k^*}$, where $k^*$ is the index of the selected fine-search codebook.
In the second step, the BS sweeps the selected fine-search codebook $\mathbf{W}^{\rm{f}}_{k^*}$. The UE also reports the corresponding power of the received signals to the BS, which is denoted as $\mathbf{z}^{\rm{f}} = \left[|y^{\rm{f}}_{k^*,1}|^2, \dots, |y^{\rm{f}}_{k^*,N_2}|^2\right]^T$ with $y^{\rm{f}}_{k^*,i} = \sqrt{\rho}\mathbf{h}^H\mathbf{w}^{\rm{f}}_{k^*,i} s + n^{\rm{f}}$. Finally, the optimal beam can be predicted by the associated beam predictor $g_{k^*}(\cdot,\diamond)$ of the fine-search codebook $\mathbf{W}_{k^*}^{\rm{f}}$ based on the measurements of both $\mathbf{z}^{\rm{c}}$ and $\mathbf{z}^{\rm{f}} $. Our task is to employ the DL techniques to jointly design the two-tier PC $\{\mathbf{W}^{\rm{c}},\{\mathbf{W}^{\rm{f}}_{k}\}_{k=1}^{G}\}$, the selector $f(\cdot)$, and the beam predictors $\{g_k(\cdot,\diamond)\}_{k=1}^{G}$ so as to minimize the beam prediction error.

Compared with the method proposed in \cite{Jeffrey}, our method adopts the hierarchical structure and hence can significantly enrich the PC space without increasing the measurement overhead for each UE.
For the multi-UE scenario, the BS may sweep more probing beams under our hierarchical method. In the worst case, the BS has to sweep all the probing beams and the number is $N_1 + GN_2$. If $G$ is not large, the increase of the sweeping overhead is tolerable.

In the following, the specific DNN architecture with a learnable two-tier PC is introduced.
\begin{figure*}[t]
\begin{centering}
\includegraphics[width=0.80 \textwidth]{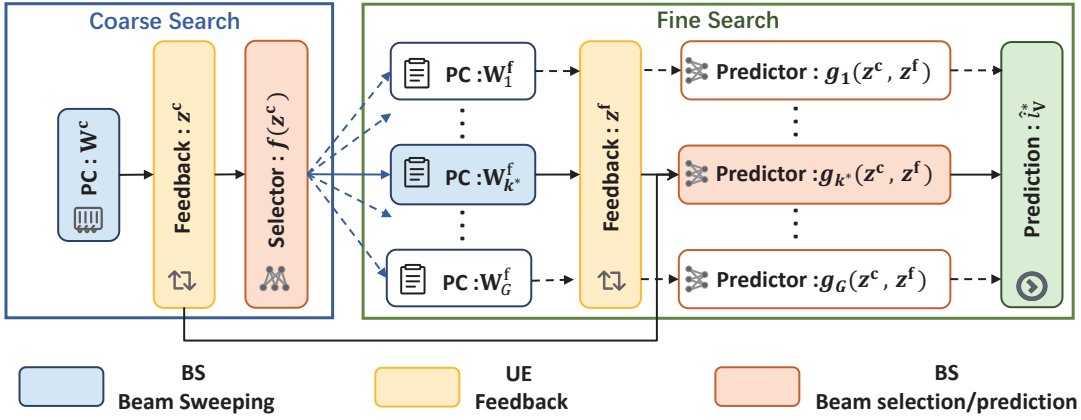}
\vspace{-0.4cm}
\caption{The diagram of the proposed hierarchical search method}\label{fig1}
\end{centering}
\vspace{-0.6cm}
\end{figure*}
\subsection{Deep Neural Network Architecture}
\begin{figure}[t]
\begin{centering}
\includegraphics[width=0.32 \textwidth]{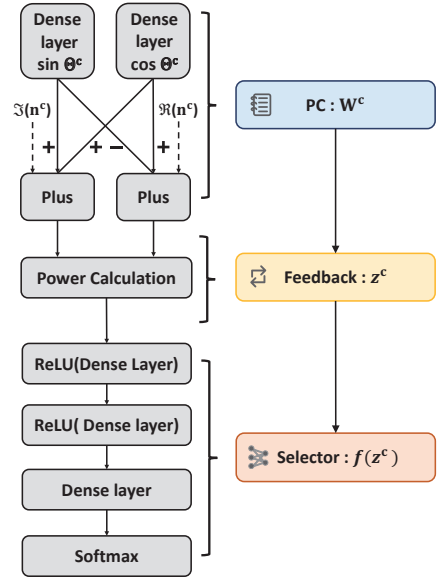}
\vspace{-0.4cm}
\caption{The DNN architecture of the coarse-search part.}\label{fig2}
\end{centering}
\vspace{-0.6cm}
\end{figure}

As shown in Fig. \ref{fig1}, the proposed hierarchical search method is realized by the DNN including a coarse-search part and a fine-search part.
In the coarse-search part, the coarse-search codebook is modeled as a complex NN layer to calculate the received signal $\mathbf{y}^{\rm{c}} = \sqrt{\rho}(\mathbf{W}^{\rm{c}})^T\mathbf{h}^{*} + \mathbf{n}^{\rm{c}} \in \mathbb{C}^{N_1 \times 1}$.
Due to the constant-modulus constraint on each element in $\mathbf{W}^{\rm{c}}$, i.e., $|w^{\rm{c}}_{i,j}| = \frac{1}{\sqrt{N_t}}$, the PC can be rewritten as
\begin{align}
    \mathbf{W}^{\rm{c}} = \frac{1}{\sqrt{N_t}}\left[\cos(\pmb{\Theta}^{\rm{c}}) + \rm{j}\cdot\sin(\pmb{\Theta}^{\rm{c}})\right],
\end{align}
where $\pmb{\Theta}^{\rm{c}} \in \mathbb{R}^{N_t \times N_1}$ is in fact the trainable parameter in the complex NN layer. As such, the calculation in the PC layer can be expressed as
\begin{align}\label{equ:z=Wh}
    \begin{bmatrix}
             \Re\{\mathbf{y}^{\rm{c}}\} \\
             \Im\{\mathbf{y}^{\rm{c}}\}
    \end{bmatrix}
    =& \sqrt{\rho}\begin{bmatrix}
             (\cos(\pmb{\Theta}^{\rm{c}}))^T&-(\sin(\pmb{\Theta}^{\rm{c}}))^T \\
             (\sin(\pmb{\Theta}^{\rm{c}}))^T&~~(\cos(\pmb{\Theta}^{\rm{c}}))^T
    \end{bmatrix}
    \begin{bmatrix}
             \Re\{\mathbf{h}^{*}\} \\
             \Im\{\mathbf{h}^{*}\}
    \end{bmatrix} \notag \\
    \quad &+ \begin{bmatrix}
             \Re\{\mathbf{n}^{\rm{c}}\} \\
             \Im\{\mathbf{n}^{\rm{c}}\}
    \end{bmatrix},
\end{align}
where the noise $\mathbf{n}^{\rm{c}}$ satisfies $\mathcal{CN}(0,\sigma^2_n\mathbf{I})$. Then the measurement $\mathbf{z}^{\rm{c}}$ is calculated by following (\ref{equ:m1}) in the feedback layer. The selector $f(\cdot)$ is realized by an multilayer perceptron (MLP) which outputs a likelihood vector $\mathbf{p} = [p_1,\dots,p_{G}]^T = f(\mathbf{z}^{\rm{c}}) \in \mathbb{R}^{G \times 1} $ to indicate the likelihood of the fine-search codebooks.
Different from the conventional hierarchical search method, here we introduce a MLP as the selector to calculate likelihood so that $N_1$ is not required to be equal to $G$.
The structure of DNN for the codebook and the operational process in the coarse-search part are shown in Fig. \ref{fig2}. The index of the selected fine-search codebook is given as
\begin{equation}\label{equ:k^*}
    k^* = \mathop{\arg\max}\limits_{k \in \{ 1,\dots,G \} } p_{k}.
\end{equation}

In the fine-search part, all the $G$ fine-search codebooks are also modeled as the complex NN layers. Similarly, we have $\mathbf{W}^{\rm{f}}_k = \frac{1}{\sqrt{N_t}}\left[\cos(\pmb{\Theta}^{\rm{f}}_k) + \mathrm{j} \cdot \sin(\pmb{\Theta}^{\rm{f}}_k)\right]$ and $\pmb{\Theta}^{\rm{f}}_k \in \mathbb{R}^{N_t \times N_2}$ is the trainable parameter in the $k$th codebook. With the selection result in (\ref{equ:k^*}), the measurement $\mathbf{z}^{\rm{f}}$ is derived through the $k^*$th fine-search PC layer and the feedback layer. The measurements $\mathbf{z}^{\rm{c}}$ and $\mathbf{z}^{\rm{f}}$ are both input to the associated beam predictor $g_{k^*}(\cdot,\diamond)$ to provide more detailed channel information and thus improve the prediction accuracy. Each beam predictor is modeled as an MLP as well. In contrast to conventional 2-tier hierarchical search methods that usually find the optimal beam from a subset of the DFT codebook $\mathbf{V}$ based on the previous searching results, the proposed beam predictor gives the likelihood $\mathbf{q}_{k^*} = [q_{k^*,1},\dots,q_{k^*,N_{\mathbf{V}}}]^T = g_{k^*}(\mathbf{z}^{\rm{c}},\mathbf{z}^{\rm{f}}) \in \mathbb{R}^{N_{\mathbf{V}} \times 1}$ of all the beams in the DFT codebook $\mathbf{V}$. In this way, the performance loss caused by decision error in the coarse-search part of conventional 2-tier hierarchical search methods can be compensated by the beam predictors in our proposed method. Finally, the index of the optimal beam is decided as
\begin{equation}\label{equ:i_V_hat}
    \hat{i}^{*}_{\mathbf{V}} = \mathop{\arg\max}\limits_{i \in \{ 1,\dots,N_{\mathbf{V}} \} } q_{k^*,i}.
\end{equation}

\subsection{Network Training}
Note that the proposed DNN is site-specific and needs to be retrained if the channel statistics change. In practice, the channel environment usually evolves slowly and remains almost static in a long period, indicating that there is no need to execute the retraining operation frequently.

Here, a dataset $\mathcal{H}$ which contains a large amount of channel vectors $\mathbf{h}$'s is adopted in the training phase.
By simulations, we find that the end-to-end training strategy usually makes the DNN converge to a bad local optimal point. Thus we propose to first train the $\mathbf{W}^{\rm{c}}$ and $f(\cdot)$. Then $\{\mathbf{W}^{\rm{f}}_{k}\}_{k=1}^{G}$ and $\{g_k(\cdot,\diamond)\}_{k=1}^{G}$ are trained based on the learned $\mathbf{W}^{\rm{c}}$ and $f(\cdot)$. Under this two-step training strategy, we find the DNN can always achieve the optimal performance.
\begin{algorithm}[h]
\begin{algorithmic}[1]
\caption{Training Procedure of the proposed DNN}\label{algorithm2}


\State \textbf{\{Training for $\{\mathbf{W}^{\rm{c}},f(\cdot)\}$\}}
\State Utilize the K-means method for channel clustering and then generate the indicator vector $\mathbf{p}^{h}$ for each channel sample;
\State Utilize $\{(\mathbf{h},\mathbf{p}^{h})\}$ as the (feature, label) pair to learn $\mathbf{W}^{\rm{c}}$ and $f(\cdot)$ with the cross-entropy function in (\ref{equ:CE_f1});

\State \textbf{\{Training for $\{\mathbf{W}^{\rm{f}}_{k}, g_k(\cdot,\diamond)\}_{k=1}^{G}$\}}
\State Select the fine-search codebook $\mathbf{W}^{\rm{f}}_{k^*}$ and the beam predictor $g_{k^*}(\cdot,\diamond)$ based on the output of  $f(\mathbf{z}^{\rm{c}})$;
\State Utilize $\{(\mathbf{h},\mathbf{q}^{\rm{h}})\}$ to learn $\mathbf{W}^{\rm{f}}_{k^*}$ and $g_{k^*}(\cdot,\diamond)$ with the cross-entropy function;

\end{algorithmic}
\end{algorithm}


\subsubsection{Training of the coarse-search codebook and the selector}
In the first training step, we train the coarse-search codebook $\mathbf{W}^{\rm{c}}$ and the selector $f(\cdot)$ jointly to make a coarse estimation of the channel.
The input is each channel sample $\mathbf{h} \in \mathcal{H}$ and the output is the likelihood vector $\mathbf{p}=\left[p_1,\dots,p_G\right]^T \in \mathbb{R}^{G \times 1}$ for the $G$ fine-search codebooks to be used. 

To facilitate the loss function design, we need to obtain the ground-truth label for each channel sample $\mathbf{h}$. Recall that if the BS employs the beams with close directions to the optimal DFT beam for a UE, such misalignment will not result in large SNR degradation. With the limited measurements, when the misalignment occurs, the predicted beam is usually close to the optimal beam. Thus a fine search concentrating on a smaller beam space can improve the performance.
Inspired by this, we propose to perform channel group by clustering channel samples with close optimal DFT beam directions.
Specifically, we first employ a DFT codebook $\mathbf{U}$ whose size $N_{\mathbf{U}}$ is much larger than $N_{\mathbf{V}}$. Then the optimal DFT beam direction of each channel sample $\mathbf{h}$ in codebook $\mathbf{U}$ is found by performing an exhaustive search. We denote the optimal beam as $\mathbf{u^*} = [1,e^{\mathrm{j}\frac{2\pi d}{\lambda}\sin(\psi)},\dots,e^{\mathrm{j}\frac{2\pi d}{\lambda}(N_t-1)\sin(\psi)}]^T$ , where $\psi$ is the optimal discrete beam direction. Note that a larger codebook size $N_{\mathbf{U}}$ can help us find a more precise beam direction of the channel sample.
Then we use the K-means clustering method to divide all the channel samples into $G$ groups based on their optimal DFT beam directions, where the distance between two channel vectors is defined as $|\sin(\psi_i) - \sin(\psi_j)|$.
After that, the group index of each channel sample is mapped to an one-hot vector $\mathbf{p}^h = [p^h_1,\dots,p^h_G]^T \in \mathbb{R}^{G \times 1}$ as the ground-truth label for training, where only $p^h_k=1$ and the other elements are all zero if the channel sample is in group $k$. 


To evaluate the difference between the predicted probability distribution and the true probability distribution, the cross-entropy function is adopted as the loss function, which can be expressed as
\begin{equation}\label{equ:CE_f1}
    L(\mathbf{p}) = -\frac{1}{G}\sum_{k=1}^{G} p^{h}_{k} \log \left( p_{k} \right).
\end{equation}


\subsubsection{Training of fine-search codebooks and beam predictors}
After the coarse-search part is well trained, the second training step is to train the fine-search codebooks $\{\mathbf{W}^{\rm{f}}_{k}\}_{k=1}^{G}$ and beam predictors $\{g_k(\cdot,\diamond)\}_{k=1}^{G}$ jointly. The input are the channel sample $\mathbf{h} \in \mathcal{H}$ and the associated output result $\mathbf{p}$ from the coarse-search part. The output is the likelihood vector $\mathbf{q}=\left[q_1,\dots,q_{N_{\mathbf{V}}}\right]^T \in \mathbb{R}^{{N_{\mathbf{V}}} \times 1}$ for the $N_{\mathbf{V}}$ beams in the DFT codebook $\mathbf{V}$.
The ground-truth label in this part is the optimal beam index in the codebook $\mathbf{V}$ and can be generated by performing an exhaustive search in $\mathbf{V}$. We also encode each label to an one-hot vector $\mathbf{q}^{\rm{h}} \in \mathbb{R}^{N_{\mathbf{V}} \times 1}$ and utilize the cross-entropy function as the loss function.
Since the data can be easily obtained in the wireless communication systems, the proposed DNN can be well trained to fully exploit the channel environment.
The whole training procedure is outlined in Algorithm \ref{algorithm2}.

\begin{table}[h]
\newcommand{\tabincell}[2]{\begin{tabular}{@{}#1@{}}#2\end{tabular}}
\caption{Probing codebook sizes in Rosslyn}\label{table1}
\vspace{-0.4cm}
\begin{center}

\begin{tabular}{|c|c|c|c|c|c|c|c|c|}
\hline
$N_1+N_2$ & 6 & 8 & 10 & 12 & 14 & 16 & 18 & 20\\
\hline
$N_1$ & 3 & 4 & 4 & 6 & 6 & 6 & 6 &  6\\
\hline
$N_2$ & 3 & 4 & 6 & 6 & 8 & 10 & 12 & 14\\
\hline
\end{tabular}
\label{tab1}
\end{center}
\vspace{-0.4cm}
\end{table}

\begin{table}[h]
\newcommand{\tabincell}[2]{\begin{tabular}{@{}#1@{}}#2\end{tabular}}
\caption{Probing codebook sizes in DeepMIMO I3}\label{table2}
\vspace{-0.4cm}
\begin{center}

\begin{tabular}{|c|c|c|c|c|c|c|c|c|}
\hline
$N_1+N_2$ & 6 & 8 & 10 & 12 & 14 & 16 & 18 & 20\\
\hline
$N_1$ & 3 & 3 & 3 & 4 & 5 & 6 & 6 &  6\\
\hline
$N_2$ & 3 & 5 & 7 & 8 & 9 & 10 & 12 & 14\\
\hline
\end{tabular}
\label{tab2}
\end{center}
\vspace{-0.4cm}
\end{table}

\section{Experiment Results}

\subsection{Dataset Description and Simulation Settings}

We perform the simulation based on the public datasets of Rosslyn \cite{ML3_rosslyn} and DeepMIMO I3 \cite{deepmimo} to compare the performance between our method and benchmarks. There are 58725 samples in Rosslyn dataset and 118959 samples in DeepMIMO I3. We use 60\% samples for training and 40\% for testing. Both the datasets are generated by ray-tracing method.
The Rosslyn scenario simulates an outdoor urban environment with few blocks while the DeepMIMO I3 scenario is indoor and several walls are placed. The carrier frequencies of the two scenarios are $28$GHz and $60$GHz, respectively.

In our simulation, the number of antenna elements in the BS is set to be $N_t=64$ and the antenna spacing is set to be $d = \frac{\lambda}{2}$. The transmit power and the noise power spectral density (PSD) in both scenarios are set to be 10 dBm and -161 dBm / Hz, respectively, if not specified otherwise. The bandwidth in these two scenarios is $100$MHz. The DFT codebook $\mathbf{V}$ for downlink transmission has $N_\mathbf{V}=128$ codewords. In addition, we set the number of fine-search codebooks $G=4$ in Rosslyn experiment and $G=3$ for DeepMIMO I3 experiment due to the different UE distributions in the two scenarios.

\subsection{Evaluating the Proposed Method}
To verify the superiority of the proposed method, four different benchmarks are evaluated for comparison, including the exhaustive beam search, binary search, 2-tier hierarchical search, and the method proposed in \cite{Jeffrey}.
The accuracy which represents the probability of correctly predicting the optimal beam is utilized as the performance metric.


\begin{figure}[t]
  \centering
  \subfigure[Rosslyn]
  {\includegraphics[width=.38\textwidth]{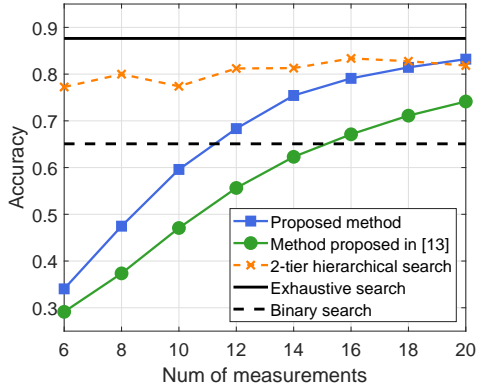}}

  \vspace{-0.3cm}
  \subfigure[DeepMIMO I3]
  {\includegraphics[width=.38\textwidth]{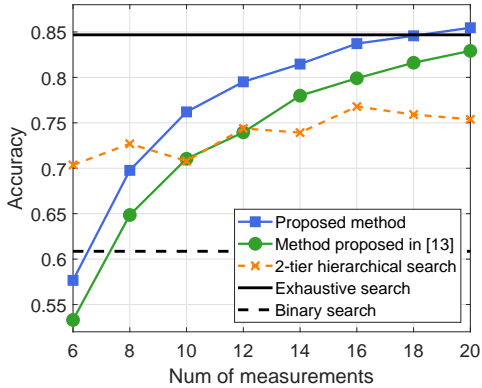}}
  \vspace{-0.3cm}
  \caption{Beam alignment accuracy v.s. Number of measurements.}\label{fig3}
  \vspace{-0.7cm}
\end{figure}

According to the results of past experiments, the sizes of codebooks in the fine-search part should be no smaller than that in the coarse-search part. This is intuitive since a more precise prediction is required in the fine-search part and more detailed channel information is necessary. The sizes of codebooks in the two parts of the proposed method are shown in Table.~\ref{table1} and Table.~\ref{table2}.
For fair comparison, we use $(N_1+N_2)$ as the number of measurements, which directly determines the sweeping overhead of the proposed beam alignment method.


Fig.~\ref{fig3} shows the accuracy of the proposed method and the considered benchmarks with respect to the number of measurements. As a remark, when $N_{\mathbf{V}}$ is fixed to 128, the exhaustive search and binary search have constant number of measurements, which are 128 and 14, respectively. For the 2-tier hierarchical search method, the BS has to sweep a narrow-beam codebook in addition to a wide-beam codebook. The figure only shows the size of its wide-beam codebook, which is set to $(N_1+N_2)$. Note that the actual number of measurements for the 2-tier hierarchical search is $\lceil N_1+N_2 + 128/(N_1+N_2)\rceil$ and consistently larger than that of the proposed method, which varies between 23$\sim $28 for all the considered $(N_1+N_2)$.
From Fig.~\ref{fig3}, we observe that when the number of measurements is small, the accuracy of our method is not impressive compared with the exhaustive search. However, as the measurement number increases, the performance of the proposed method is greatly improved. This indicates that the larger PC can help the BS collect more detailed information about the channel.
When the number of measurements is set to be 20 in DeepMIMO I3 scenario, the proposed method outperforms all benchmarks including the exhaustive search, this is due to that the DL methods can also learn the statistical property of noise, which is lacked in conventional methods.
It is also noted that our method can consistently achieve a better performance than the method proposed in \cite{Jeffrey}. In particular, when $N_1+N_2=14$, the accuracy of our method is 12.8\% and 5.1\% higher than that of the method proposed in \cite{Jeffrey} in these two scenarios, respectively.
In addition, we find that the DL based methods can achieve a much better performance than conventional  hierarchical search methods in the DeepMIMO I3 scenario. This suggests that the DL methods can significantly be adapt to different scenarios.

\begin{figure}[t]
  \centering
  \subfigure[Rosslyn]
  {\includegraphics[width=.38\textwidth]{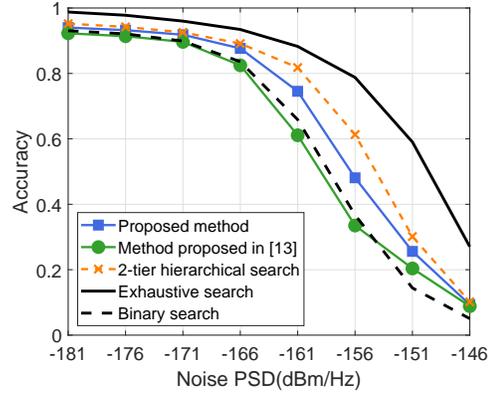}}

  \vspace{-0.3cm}
  \subfigure[DeepMIMO I3]
  {\includegraphics[width=.38\textwidth]{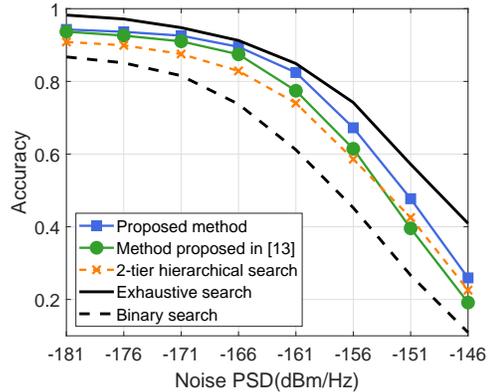}}
  \vspace{-0.3cm}
  \caption{Beam alignment accuracy v.s. Noise PSD.}\label{fig4}
  \vspace{-0.7cm}
\end{figure}

\begin{figure*}[t]
  \centering
  \subfigure[$\mathbf{W}^{\rm{c}}$]
  {\includegraphics[width=.18\textwidth]{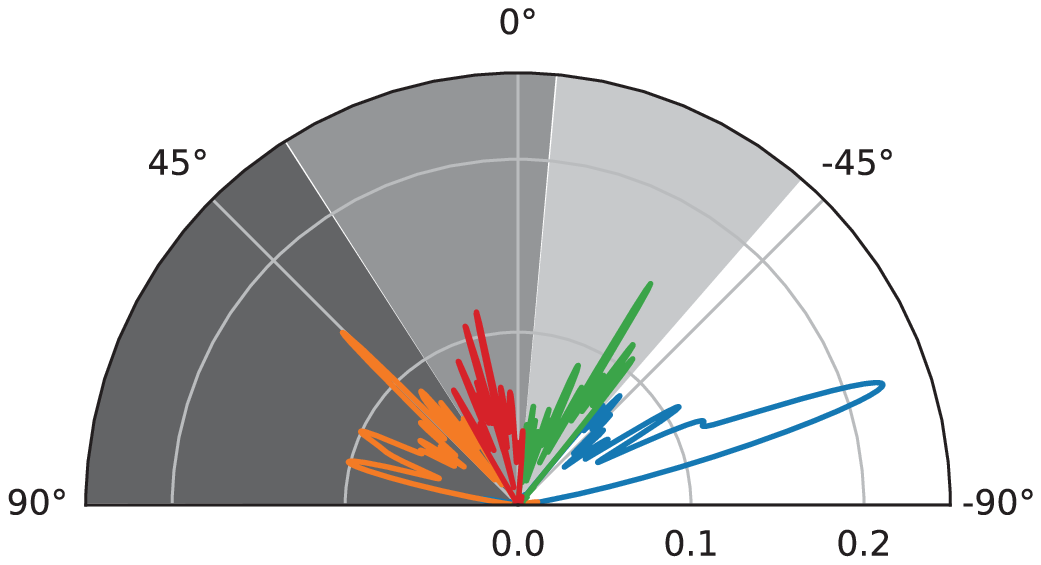}\label{fig5a}}
  \subfigure[$\mathbf{W}^{\rm{f}}_{1}$]
  {\includegraphics[width=.18\textwidth]{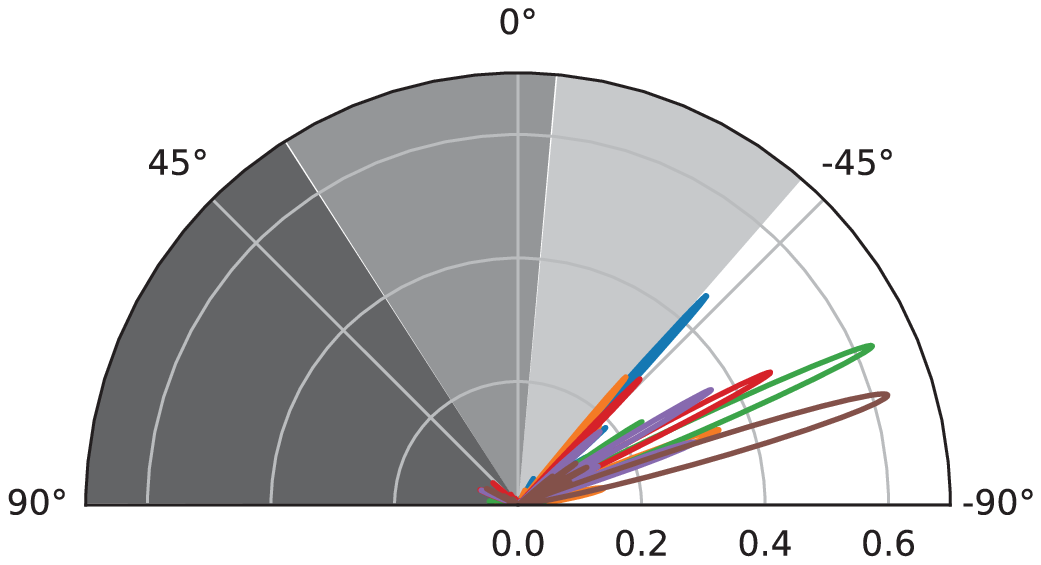}\label{fig5b}}
  \subfigure[$\mathbf{W}^{\rm{f}}_{2}$]
  {\includegraphics[width=.18\textwidth]{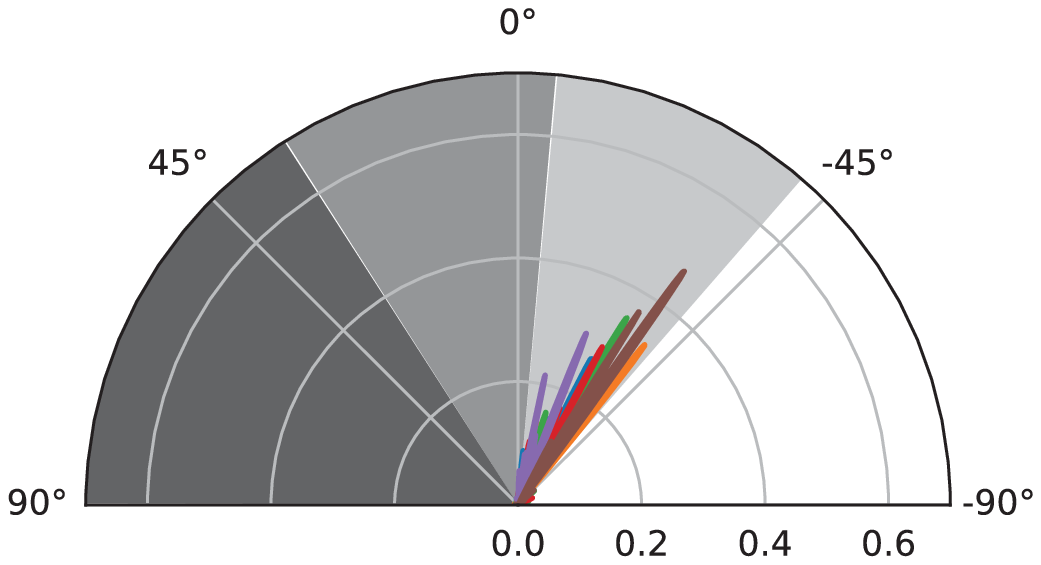}\label{fig5c}}
  \subfigure[$\mathbf{W}^{\rm{f}}_{3}$]
  {\includegraphics[width=.18\textwidth]{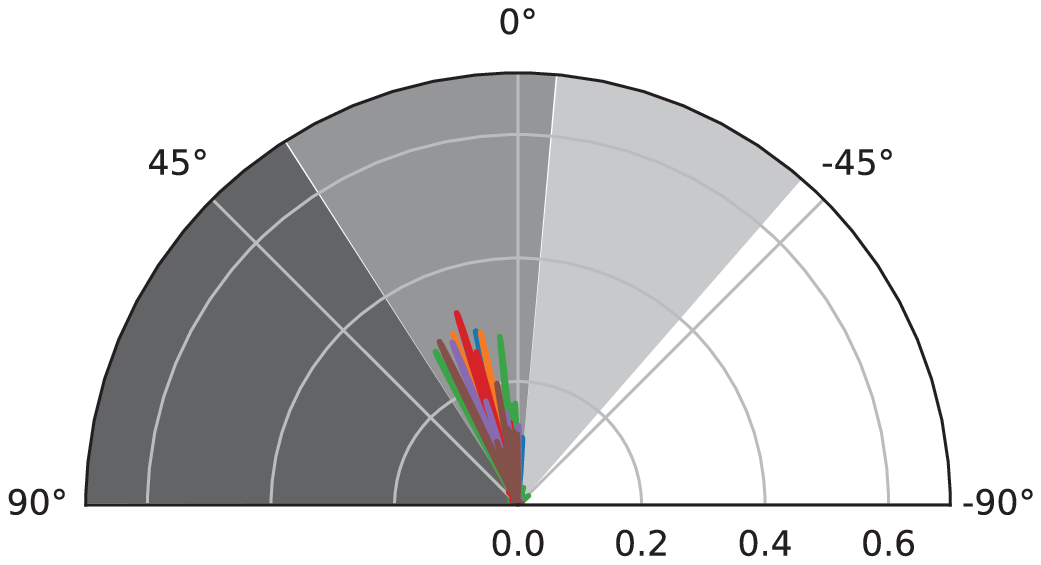}\label{fig5d}}
  \subfigure[$\mathbf{W}^{\rm{f}}_{4}$]
  {\includegraphics[width=.18\textwidth]{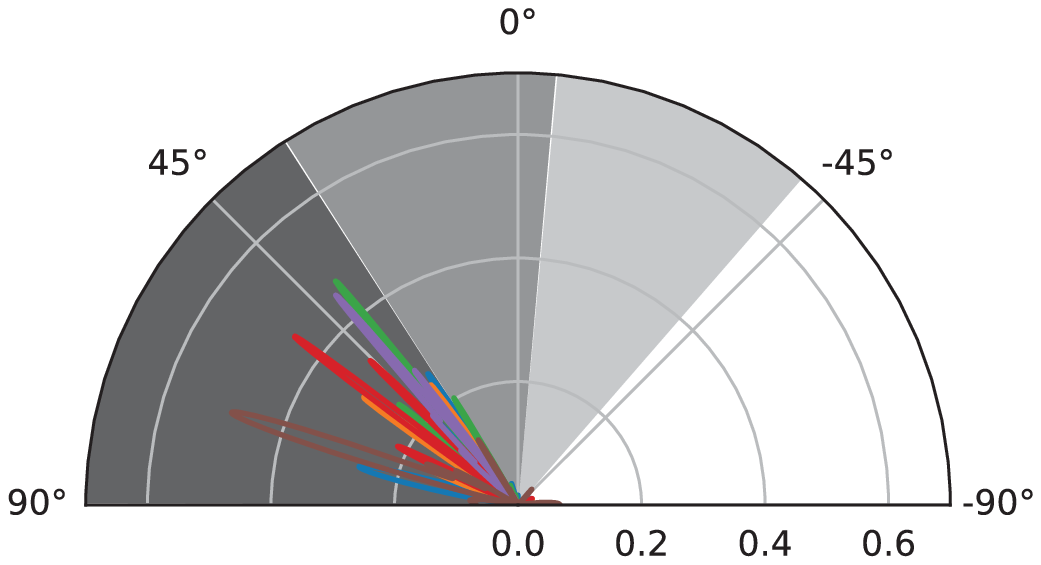}\label{fig5e}}

  \vspace{-0.3cm}
  \caption{The beam pattern of the learned two-tier PC in Rosslyn.}\label{fig5}
  \vspace{-0.7cm}
\end{figure*}

To examine the robustness of the proposed methods, we exam the accuracy of these methods with different noise PSD. In the simulation, the number of measurements is set to be $N_1 + N_2 = 14$ for fair comparison with the binary search. According to Fig.~\ref{fig4}, the performance of all methods are gradually deteriorated as the noise PSD increases. When the noise PSD is smaller than -166 dBm/Hz, the performance of our method is similar to or even better than the 2-tier hierarchical search method with only 60.5\% sweeping overhead of it.
In addition, it can be observed that our method can always outperform the method proposed in \cite{Jeffrey} in different noise PSD. For example, when the noise PSD is -156 dBm/Hz, the proposed method improves the accuracy by 14.6\% in Rosslyn experiment and by 6.2\% in DeepMIMO I3 experiment.

\subsection{Learned PC patterns}
The two-tier PC is designed to collect channel information for the site-specific BS. The Fig.~\ref{fig5} shows the PC trained in Rosslyn experiment, the Fig.~\ref{fig5a} is the pattern of the coarse-search codebook and the rest shows the fine-search codebook. We set $N_1=4$ and $N_2=6$, respectively. Different grey levels are utilized to represent the groups generated by the K-means method, i.e., labels in the coarse-search part. It is noted that the pattern of coarse-search codebook $\mathbf{W}^{\rm{c}}$ exactly fits to the clustering results. Similar to the pattern of codebook $\mathbf{W}^{\rm{c}}$, each codebook in the fine-search part is learned to focus on searching a particular region.

\begin{figure}[t] 

\centering
\subfigure[$\mathbf{W}^{\rm{c}}$]
{\includegraphics[width=.18\textwidth]{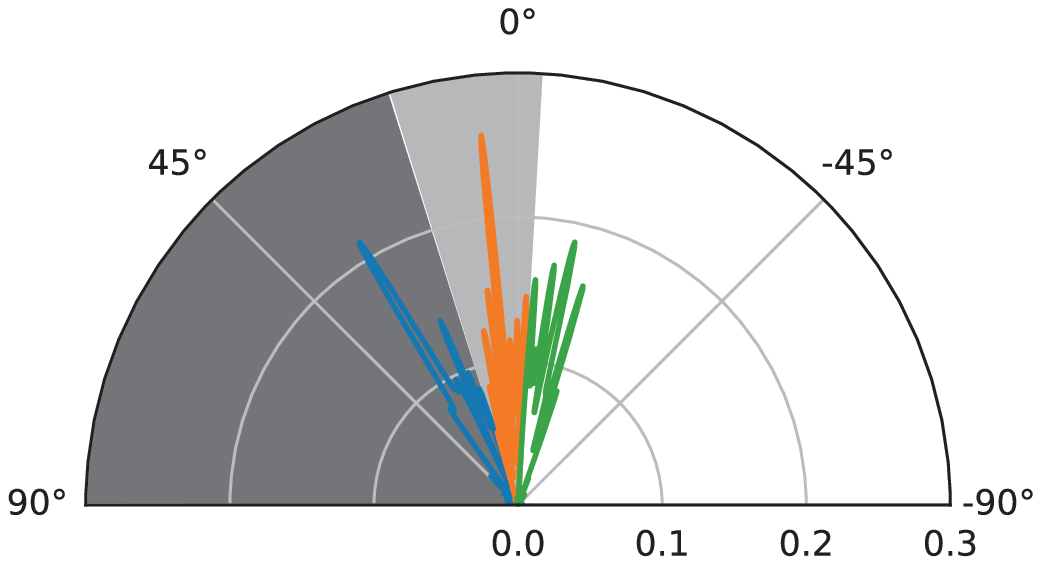}\label{fig6a}}
\subfigure[$\mathbf{W}^{\rm{f}}_{1}$]
{\includegraphics[width=.18\textwidth]{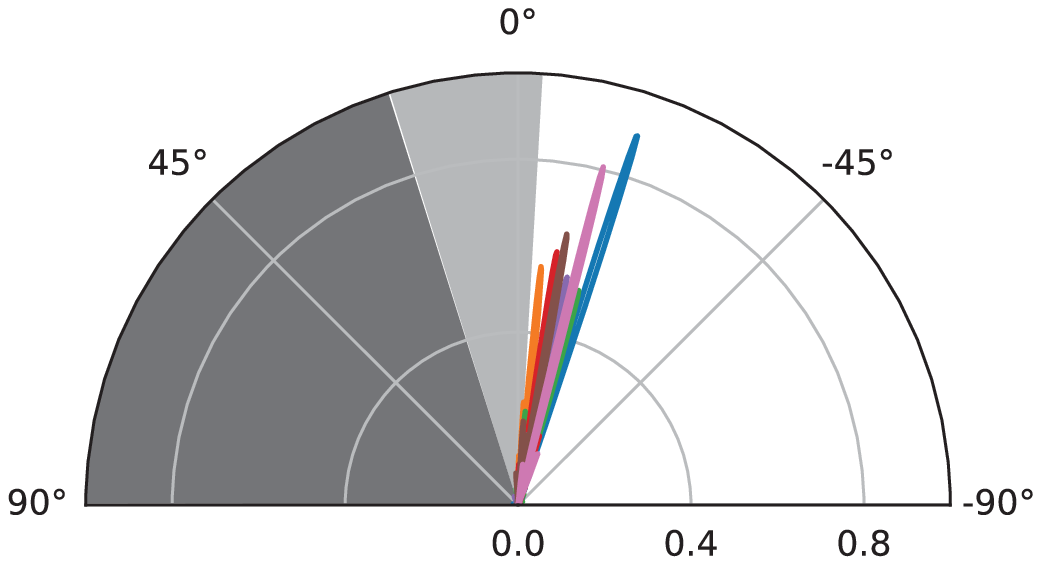}\label{fig6b}}
\subfigure[$\mathbf{W}^{\rm{f}}_{2}$]
{\includegraphics[width=.18\textwidth]{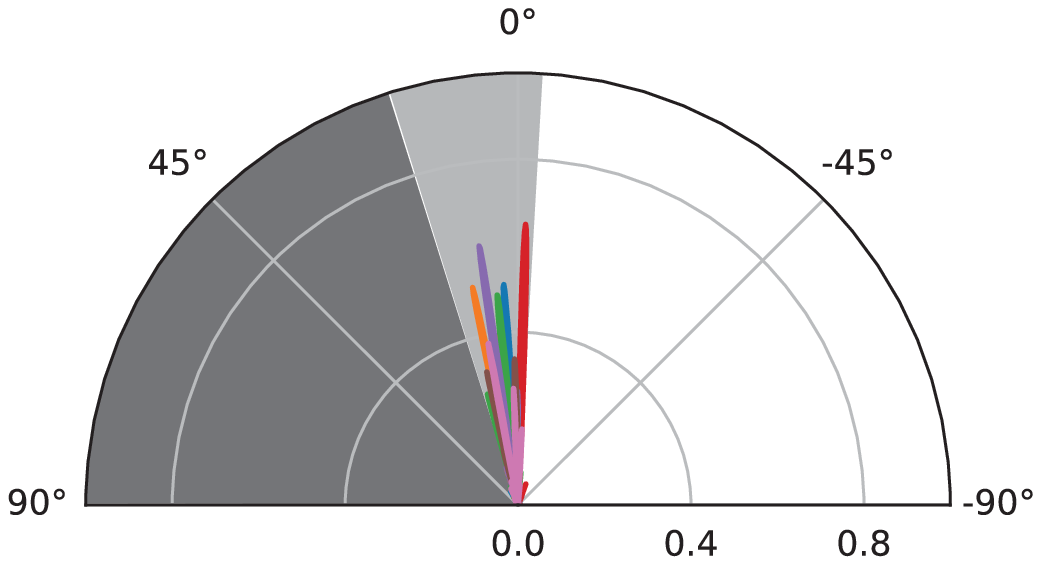}\label{fig6c}}
\subfigure[$\mathbf{W}^{\rm{f}}_{3}$]
{\includegraphics[width=.18\textwidth]{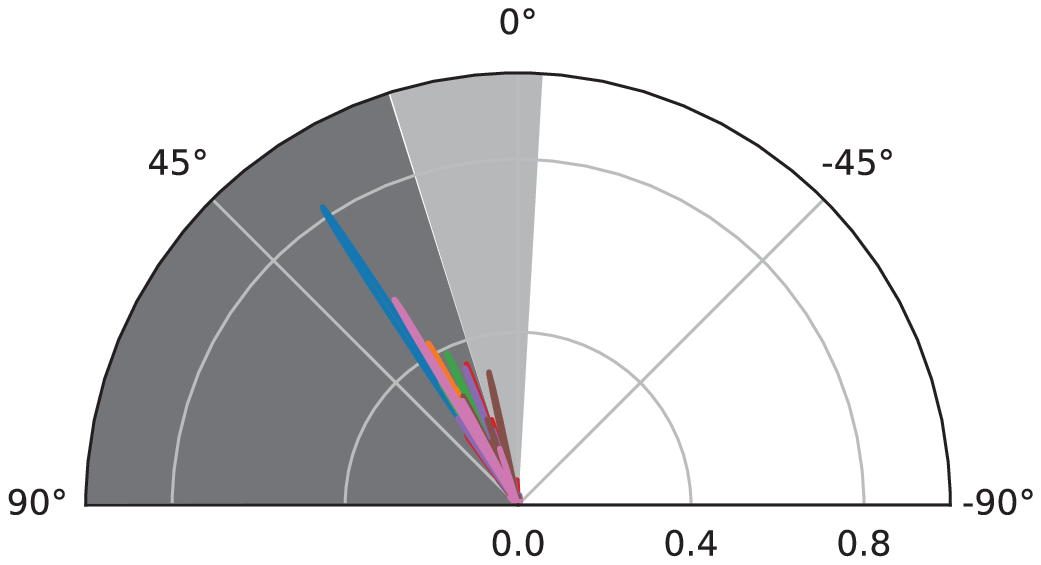}\label{fig6d}}
\vspace{-0.2cm}
\caption{The beam pattern of the learned two-tier PC in DeepMIMO I3.}\label{fig6}
\vspace{-0.7cm}
\end{figure}

The PC for DeepMIMO I3 scenario is shown in Fig.~\ref{fig6}. The codebook sizes are $N_1=3$ and $N_2=7$, respectively.
UEs in the DeepMIMO I3 scenario are mainly concentrated in the front of the BS and it is clearly that comparing to the probing beam patterns trained by Rosslyn dataset, the beams in DeepMIMO I3 scenario are much more focused. This indicates that the proposed DNN structure can be efficiently trained to be adapt to different environments.

\section{Conclusion}
In this paper, we investigate the beam alignment for mmWave communication system. 
A novel DL-based method with a learned hierarchical PC is proposed to predict the optimal beam in a coarse-to-fine search way. We also propose an effective training strategy which is performed in two steps for the proposed method. Simulation results show the superior performance of our method comparing to other existing alternatives and the two-tier PC can be learned to effectively capture the features of the propagation environment.

\bibliographystyle{IEEEtran}
\bibliography{refer}

\begin{thebibliography}{10}
\providecommand{\url}[1]{#1}
\csname url@samestyle\endcsname
\providecommand{\newblock}{\relax}
\providecommand{\bibinfo}[2]{#2}
\providecommand{\BIBentrySTDinterwordspacing}{\spaceskip=0pt\relax}
\providecommand{\BIBentryALTinterwordstretchfactor}{4}
\providecommand{\BIBentryALTinterwordspacing}{\spaceskip=\fontdimen2\font plus
\BIBentryALTinterwordstretchfactor\fontdimen3\font minus
  \fontdimen4\font\relax}
\providecommand{\BIBforeignlanguage}[2]{{%
\expandafter\ifx\csname l@#1\endcsname\relax
\typeout{** WARNING: IEEEtran.bst: No hyphenation pattern has been}%
\typeout{** loaded for the language `#1'. Using the pattern for}%
\typeout{** the default language instead.}%
\else
\language=\csname l@#1\endcsname
\fi
#2}}
\providecommand{\BIBdecl}{\relax}
\BIBdecl

\bibitem{background1_hierarchical1}
A.~Alkhateeb, O.~El~Ayach, G.~Leus, and R.~W. Heath, ``Channel estimation and
  hybrid precoding for millimeter wave cellular systems,'' \emph{IEEE Journal
  of Selected Topics in Signal Processing}, vol.~8, no.~5, pp. 831--846, 2014.

\bibitem{background2}
S.~Kutty and D.~Sen, ``Beamforming for millimeter wave communications: An
  inclusive survey,'' \emph{IEEE Communications Surveys Tutorials}, vol.~18,
  no.~2, pp. 949--973, 2016.

\bibitem{exhaustive1}
C.~Jeong, J.~Park, and H.~Yu, ``Random access in millimeter-wave beamforming
  cellular networks: issues and approaches,'' \emph{IEEE Communications
  Magazine}, vol.~53, no.~1, pp. 180--185, 2015.

\bibitem{exhaustive2}
C.~N. Barati, S.~A. Hosseini, S.~Rangan, P.~Liu, T.~Korakis, S.~S. Panwar, and
  T.~S. Rappaport, ``Directional cell discovery in millimeter wave cellular
  networks,'' \emph{IEEE Transactions on Wireless Communications}, vol.~14,
  no.~12, pp. 6664--6678, 2015.

\bibitem{hierarchical2}
Z.~Xiao, T.~He, P.~Xia, and X.-G. Xia, ``Hierarchical codebook design for
  beamforming training in millimeter-wave communication,'' \emph{IEEE
  Transactions on Wireless Communications}, vol.~15, no.~5, pp. 3380--3392,
  2016.

\bibitem{hierarchical3}
C.~Qi, K.~Chen, O.~A. Dobre, and G.~Y. Li, ``Hierarchical codebook-based
  multiuser beam training for millimeter wave massive mimo,'' \emph{IEEE
  Transactions on Wireless Communications}, vol.~19, no.~12, pp. 8142--8152,
  2020.

\bibitem{ML1}
Y.~Wang, M.~Narasimha, and R.~W. Heath, ``Mmwave beam prediction with
  situational awareness: A machine learning approach,'' in \emph{2018 IEEE 19th
  International Workshop on Signal Processing Advances in Wireless
  Communications (SPAWC)}, 2018, pp. 1--5.

\bibitem{ML2}
C.~Antón-Haro and X.~Mestre, ``Learning and data-driven beam selection for
  mmwave communications: An angle of arrival-based approach,'' \emph{IEEE
  Access}, vol.~7, pp. 20\,404--20\,415, 2019.

\bibitem{ML3_rosslyn}
Y.~Heng and J.~G. Andrews, ``Machine learning-assisted beam alignment for
  mmwave systems,'' \emph{IEEE Transactions on Cognitive Communications and
  Networking}, vol.~7, no.~4, pp. 1142--1155, 2021.

\bibitem{reinforcement1}
F.~B. Mismar, B.~L. Evans, and A.~Alkhateeb, ``Deep reinforcement learning for
  5g networks: Joint beamforming, power control, and interference
  coordination,'' \emph{IEEE Transactions on Communications}, vol.~68, no.~3,
  pp. 1581--1592, 2020.

\bibitem{reinforcement2}
Y.~Zhang, M.~Alrabeiah, and A.~Alkhateeb, ``Reinforcement learning of beam
  codebooks in millimeter wave and terahertz mimo systems,'' \emph{IEEE
  Transactions on Communications}, vol.~70, no.~2, pp. 904--919, 2022.

\bibitem{reinforcement3}
\BIBentryALTinterwordspacing
V.~Raj, N.~Nayak, and S.~Kalyani, ``Deep reinforcement learning based blind
  mmwave mimo beam alignment,'' 2020. [Online]. Available:
  \url{https://arxiv.org/abs/2001.09251}
\BIBentrySTDinterwordspacing

\bibitem{Jeffrey}
Y.~Heng, J.~Mo, and J.~G. Andrews, ``Learning site-specific probing beams for
  fast mmwave beam alignment,'' \emph{IEEE Transactions on Wireless
  Communications}, pp. 1--1, 2022.

\bibitem{software}
\emph{Wireless InSite 3.2.0 Reference Manual}, Remcom Inc., 2017. [Online].
  Available: https://www.remcom.com/wireless-insite-em-propagation-software/.

\bibitem{deepmimo}
\BIBentryALTinterwordspacing
A.~Alkhateeb, ``Deepmimo: A generic deep learning dataset for millimeter wave
  and massive mimo applications,'' 2019. [Online]. Available:
  \url{https://arxiv.org/abs/1902.06435}
\BIBentrySTDinterwordspacing

\end{thebibliography}
	
\end{document}